\input epsf
\documentstyle[aps]{revtex}

\begin{document}
\twocolumn[\hsize\textwidth\columnwidth\hsize\csname@twocolumnfalse\endcsname
\draft
\title{ Superfluid anisotropy in YBCO: 
Evidence for pair tunneling superconductivity}
 
\author{T. Xiang and J. M. Wheatley}
\address{ Research Center in Superconductivity, 
University of Cambridge, Madingley Road, \\
Cambridge CB3 0HE, United Kingdom}
\date{\today}
\maketitle
\begin{abstract}
Proximity effect and pair tunneling models are applied as alternative 
scenarios to explain the recently measured $ab$-plane and $c$-axis components 
of the superfluid tensor in 
Copper-Oxide superconductors which contain chains, 
such as $YBa_2Cu_3O_{6.95}$. It is argued that
conventional proximity effect models, which couple 
chains and planes via single electron tunneling only,
are incompatible with the experimental observations. 
On the other hand several surprising features of the 
experimental data are readily explained by the presence of
a microscopic pair tunneling process.
\end{abstract}
]

\noindent\underline{\it Introduction}
Recent measurements of the electromagnetic response of 
clean untwinned single crystals of $YBa_2Cu_3O_{7-\delta}$ 
and $YBa_2Cu_4O_8$ reveal a large $ab$-plane 
anisotropy\cite{zhang,tallon}. The ratio of the superfluid 
density in $b$ and $a$ directions (CuO chains are along the $b$-axis) 
at zero temperature is about $\approx 2.4$ for the 
one-chain compound $YBa_2Cu_3O_{6.95}$ and rather larger, 
$\approx 6$ for the two-chain compound 
$YBa_2Cu_4O_8$\cite{zhang}. Anisotropies of similar 
magnitudes are observed in the normal state resistivity\cite{friedmann}. 
Early work on Knight shift and the NMR relaxation 
time $T_1$\cite{takigawa} in $YBa_2Cu_3O_{6.95}$, 
indicates that there is an  appreciable gap on the 
chains below $T_c$. Taken together, these experiments 
unequivocally show that normal electrons on the 
CuO chains are coupled into the superconducting state below $T_c$. 
However, an unexpected feature of the recent data is that 
the temperature dependences of the superfluid densities 
in $a$ and $b$ directions are similar; both
are clearly linear at low temperature\cite{zhang}, 
and roughly obey $\rho_a^{(s)}(T)/\rho_a^{(s)}(0) \simeq 
\rho_b^{(s)}(T)/\rho_b^{(s)}(0)$ up to $T_c$,
where $\rho_{a(b)}^{(s)}$ is the superfluid density 
in the $a(b)$ direction. Along the c-axis, $\rho_c^{(s)}(T) $
approaches its zero temperature value with a power higher 
than linear. The implications of these facts for the microscopic
nature of c-axis coupling in Copper-Oxides is 
the subject of this paper.
 
Two distinct models of plane-chain coupling are considered here;  
firstly a proximity model where intrinsically superconducting 
plane and chain layers are coupled through {\it single electron 
tunneling only}\cite{klemm} and secondly, an interlayer pair tunneling model 
where chains and planes are coupled through a Josephson-like 
{\it pair tunneling} process. In both cases the superfluid response 
is assumed to be dominated by coherent Fermi excitations about
quasi-2D chain and plane Fermi surface sheets\cite{arpes}. 
We mimic clean $YBCO$-type electronic structures by stacking 
planar ``$CuO_2$" and chain ``$CuO$" layers 
alternately along the c-axis, and employ a simple tight-binding description.
We show that the penetration depth results 
are incompatible with the proximity model, but find a natural explanation
within the pair tunneling model. 

\noindent\underline{\it Proximity model:} 
The proximity model is defined by the mean-field 
Hamiltonian  
\begin{eqnarray}
H&=&H_0+H_1 \label{ham}
\\
H_0&=& \sum_{k} \left\{ \sum_n \varepsilon_{n k}c_{nk\sigma}^\dagger 
c_{nk\sigma} +\varepsilon_{\perp  k} (c_{1k\sigma}^\dagger 
c_{2k\sigma} + h.c.)\right\} \label{h0}
\\ 
H_1&=& \sum_{nk}\Delta_n\gamma_{nk} (c_{nk\uparrow}^\dagger 
c_{n -k\downarrow}^\dagger + {\rm h.c.}) \label{h1}
\end{eqnarray}
where $c^\dagger_{1k\sigma}$ and $c^\dagger_{2k\sigma}$ are the 
creation operators of electrons in the plane and chain bands 
respectively, $\varepsilon_{1\, k}=- 2t(\cos k_a+\cos k_b)-\mu$, 
and $\varepsilon_{2\, k} =-2 t_c \cos k_b -\mu +\varepsilon_c$. 
Here $\mu$ the chemical potential and 
$\varepsilon_c$ the relative energy between the chains and the planes.  
It is sufficient to assume the simplest nearest neighbor 
tight binding dispersions for electrons on both chain and plane layers; 
the qualitative results are independent of the detailed forms of 
$\varepsilon_{1k}$ and $\varepsilon_{2k}$. $\Delta_n\gamma_{nk}$ are 
the gap functions with the pairing symmetries $\gamma_{nk}$ on the 
plane ($n=1$) and chain ($n=2$) bands. They are determined by 
self-consistent gap equations 
${\lambda_n / V}\sum_k \gamma_{nk} \langle c_{nk\uparrow}
c_{n\,-k\downarrow}\rangle =\Delta_{n}$, 
where the $\lambda_n$ ($n$=1, 2) are the strengths of the pairing 
potential on the plane and chain layers. 
When $\lambda_2=0$, the chain layer is not intrinsically superconducting, 
$\Delta_2=0$, and the superconducting correlation in the chain 
layer is generated purely by the proximity effect. 
The coupling between the chains and planes in Eq.~\ref{h0}
is via single electron interlayer hopping with matrix 
element $\varepsilon_{\perp \, k} = -2 t_\perp \cos(k_c/2)$. 
This coupling vanishes at the zone boundary $k_c = \pi$.
In the physically relevant regime the interlayer 
hopping constant $t_\perp$ is much smaller 
than the plane and chain hopping constants $t$ and $t_c$. 

The mean-field Hamiltonian (\ref{ham}) has four branches of 
quasiparticle excitations. 
Their energy dispersions are given by $\pm E_{\pm ,k}$ with 
$E_{\pm ,k}= \sqrt{b_k\pm\sqrt{b_k^2-c_k}}$,
where $b_k=\varepsilon_{\perp k}^2+\sum_n( \Delta_n^2\gamma_{nk}^2
+\varepsilon^2_{nk})/2$, and $c_k=(\varepsilon_{1k} \Delta_2\gamma_{2k}+
\varepsilon_{2k} \Delta_1\gamma_{1k})^2+(\varepsilon_{\perp k}^2-
\varepsilon_{1k}\varepsilon_{2k}+ \Delta_1\Delta_2\gamma_{1k}
 \gamma_{2k})^2$. 
While the $E_{+,k}$ quasiparticle excitation branch has a gap for all $k$,
zero energy excitations may be present in the $E_{-,k}$ 
branch. 
The locations of such energy gap nodes are given by 
the condition $c_k=0$, i.e. by the solutions of the simultaneous equations
$\varepsilon_{1k} \Delta_2\gamma_{2k}+
\varepsilon_{2k} \Delta_1\gamma_{1k}=0$ and $\varepsilon_{\perp k}^2-
\varepsilon_{1k}\varepsilon_{2k}+ 
\Delta_1\Delta_2\gamma_{1k}\gamma_{2k}=0$. 
Such solutions depend strongly on the pairing 
symmetries of electrons on both layers. 
If the planar electrons have d-wave pairing symmetry, i.e. 
$\gamma_{1k}=\cos k_b-\cos k_a$, line gap nodes 
always exist. However, if both $\gamma_{1k}$ and  $\gamma_{2k}$ have
isotropic s-wave symmetry (e.g. $\gamma_{1k}$ =
$\gamma_{2k}$=1) nodal lines exist only 
when the pairing functions on the plane and 
chain layers have opposite signs, $ \Delta_{1}$$ \Delta_{2}
\le 0$, and, in addition, $4t_\perp^2 > | \Delta_{1}| | \Delta_{2}|$. 
In the somewhat artificial case where the pair interaction 
on the chain layer vanishes, 
$\Delta_{2}=0$ and the ``proximity'' nodal line on the chain
Fermi surface sheet ($\varepsilon_{2k}=0$) lies along $k_c=\pi$, 
independent of the pairing symmetry of the plane layer. 
More generally, proximity induced 
nodal line loops are present on the chain Fermi surface sheet. 

\begin{figure}
\leavevmode
\epsfxsize=8.5cm
\epsfbox{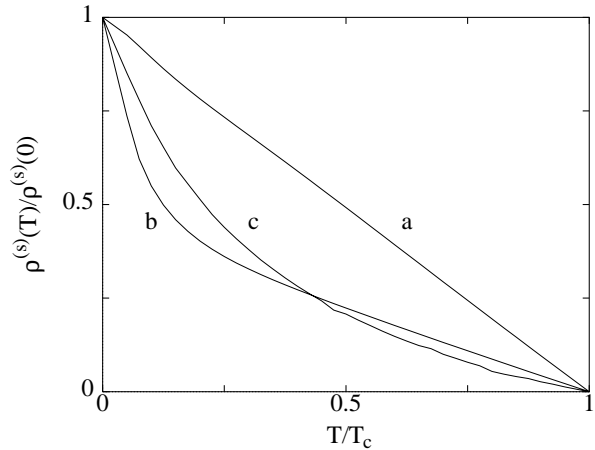}
\caption{Normalized superfluid densities $\rho^{(s)}(T)/ \rho^{(s)}(0)$ 
as functions of $T/T_c$ along $a$, $b$, and $c$ directions 
in the proximity model. 
$\gamma_{1k}=\cos k_a -\cos k_b$ and $\gamma_{2k}=\cos k_b$. 
$t=1$, $t_c=0.85$, $t_\perp =0.17$, $\varepsilon_c=0.5$, 
$\lambda_1 =\lambda_2=1.5$ 
($T_c\sim 0.23$), and the electron concentration on the 
CuO planes is about 0.85. 
\label{fig1}}
\end{figure}

The superfluid tensor is related to the kinetic energy and the 
current-current correlation function\cite{scalapino}.  
For the proximity model Eq.~\ref{ham}, 
we find that the modification of the 
superfluid density along the non-chain $a$-axis ($\rho^{(s)}_a$)
due to chain-plane hybridization is 
very small, and its temperature dependence is 
almost the same as for a pure 2D system with the chains and planes 
decoupled. If the planar band has d-wave pairing 
symmetry, $\rho_a^{(s)}$ will be linear in $T$ at low temperature. 
Along the $b$-axis, however, 
the contribution of the chain band is important. 
Close to the transition temperature, 
$\rho^{(s)}_b$ is dominated by the contribution 
of the plane band and the difference between $\rho^{(s)}_a$ and 
$\rho^{(s)}_b$ is small. At low temperature, the 
leading temperature dependence of $\rho^{(s)}_b$ is determined by that of
the low-energy density of states $\rho (\omega ) 
= {1/V}\sum_k \{ \delta (\omega -E_{-,k})+\delta (\omega -E_{+,k})\}$. 
In the absence of nodes, for example when $\gamma_{1k}=\gamma_{2k}=1$ 
and $\Delta_1\Delta_2 > 0$, $\rho^{(s)}_b$ approaches to its 
zero-temperature value exponentially as the temperature decreases. 
Since this is clearly inconsistent with 
experiments, we consider only (nodeful)
gapless cases. When $\omega < \min (E_{+,k})$, only $E_{-,k}$ 
quasiparticles have contributions to $\rho (\omega )$. Since 
$E_{-,k}=\sqrt{c_k}/E_{+,k}$ and $E_{+,k}$ is weakly k-dependent 
around the gap nodes in general, $\rho (\omega )$ 
at low energy is therefore determined mainly by 
the structure of $c_k$. If $ \Delta_2 =0 $,  
the quasiparticles around the nodal line $k_c=\pi$ and 
$\varepsilon_{2k}=0$ dominate the low energy excitations, and it can 
be shown that $\rho (\omega )\sim \sqrt{\omega}$ 
for small $\omega$. This is a 
consequence of the extreme flatness of the quasi-particle 
dispersion in a direction
normal to the nodal line $E_{-,k} \sim (k_c - \pi)^2$ near $k_c=\pi$. 
Correspondingly, the low temperature superfluid density 
along the $b$-direction has
pronounced upward curvature,
increasing rapidly as $\sqrt{T}$ as zero temperature is 
approached\cite{atkinson}. This behavior is clearly
inconsistent with the linear temperature 
dependence of the superfluid density in both planar 
directions in $YBa_2Cu_3O_{7-\delta}$ and $YBa_2Cu_4O_8$.
On the other hand, in the realistic situation where both 
$\Delta_1$ and $\Delta_2$ are finite, 
it can be shown that $\rho(\omega )\sim \omega$ when $\omega
\ll (|\Delta_1|$, $|\Delta_2|)$, and hence $\rho^{(s)}_b\sim T$ 
when $T\ll (|\Delta_1|$, $|\Delta_2|)$. 
In an intermediate temperature regime 
$|\Delta_2|\ll T\ll |\Delta_1|$, the strong upward
curvature in $\rho^{(s)}_b$ persists. 
Moreover, in the regime of physically relevant parameters  
self-consistent solutions of the gap equations 
give $|\Delta_2|\ll |\Delta_1|$. 
Along the $c$-axis, $\rho_c^{(s)}$ shows also a positive curvature.  
But the low temperature behavior of 
$\rho_c^{(s)}$ is more peculiar. If $\Delta_2=0$, 
it can be shown that at low temperature 
$\rho_c^{(s)}\sim T$ rather than $\sqrt{T}$ 
as a consequence of vanishing coherence factors on proximity nodal lines. 
If, however, $\Delta_2$ is finite and $T \ll \Delta_2$, we find that 
$\rho_c^{(s)}\sim T^2$. 
Figure \ref{fig1} shows the typical temperature dependence 
of the three components of the superfluid 
densities in the proximity model on a normalized plot. 

\begin{figure}
\leavevmode
\epsfxsize=8.5cm
\epsfbox{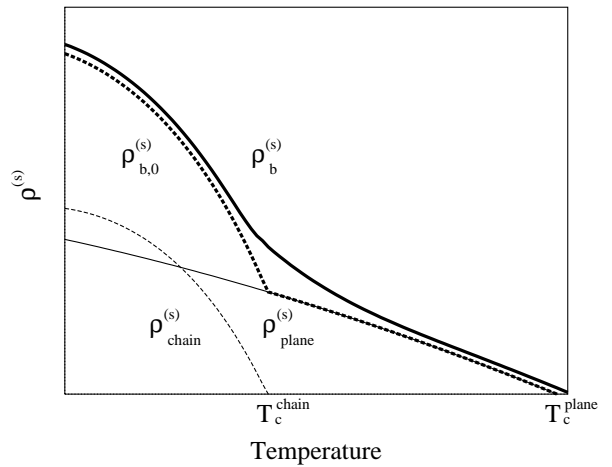}
\caption{Schematic representation of the superfluid density 
in the proximity model. $\rho^{(s)}_{chain}$ and  
$\rho^{(s)}_{plane}$ are the superfluid densities 
of the chains and planes when the interlayer coupling is zero. 
$\rho_{b,0}^{(s)}=\rho^{(s)}_{plane}+\rho^{(s)}_{chain}$. When
chains and planes are weakly coupled by single electron tunneling,
the chain-direction superfluid response develops a
positive curvature near $T_{c}^{chain}$ .
\label{fig2}}
\end{figure}

The presence of a positive curvature in $\rho_b^{(s)}$ is 
not limited to the ``pure" proximity effect limit
discussed above but is in fact
a general feature in weakly coupled two-gap 
systems. Consider the 
case where the chain and plane bands are completely 
decoupled with transition temperatures 
$T^{\rm chain}_c < T^{\rm plane}_c$ and
superfluid densities $\rho^{(s)}_{\rm chain}$ and 
$\rho^{(s)}_{\rm plane}$, respectively.  
In this case, $\rho^{(s)}_b= 
\rho^{(s)}_{\rm plane}+\rho^{(s)}_{\rm chain}$ will 
have a sudden change at $T_c^{\rm chain}$ as shown in Figure \ref{fig2}. 
Switching on a weak interlayer coupling $t_{\perp}$,
leads to  single transition temperature $\simeq T_{c}^{\rm plane}$ but leaves 
a smooth upturn in $\rho_b^{(s)}$ in the vicinity of  $T_c^{\rm chain}$.

While a $d_{x^2-y^2}$ paired state can account for the linear temperature 
behavior of $\rho_a^{(s)}$, we conclude that the proximity model
cannot give a satisfactory explanation of the observed 
temperature dependences of $\rho_b^{(s)}$ or $\rho_c^{(s)}$. 
However, the above analysis does highlight 
paradoxical features of the experimental results, namely,    
the CuO chain layers must be intrinsically superconducting, 
yet there must be {\it a node of the energy gap on the 
chain Fermi surface sheet}. Otherwise, 
${({\rm d} \rho_b^{(s)} / {\rm d} T)}|_{T=0}
\simeq {({\rm d} \rho_a^{(s)}/ {\rm d} T)}|_{T=0}$ as the chain 
band gives no contribution to the linear temperature term of 
$\rho_b^{(s)}$ at low temperature, in contradiction to 
experiment. As a quasi-1D system has an energy gap under 
ordinary circumstances, the presence of nodes suggests that the 
gap function of the chain band $\Delta_2\gamma_{2k}$ may
have 2D character; i.e. $\gamma_{2k}$ is not simply 
a function of $k_b$ only. We are thus led to the surprising conclusion
that superconducting coherence extends
across CuO chains in the chain layer.

\noindent\underline{\it Interlayer pair tunneling model:} 
The failure of the single particle tunneling model
leads us to introduce a pair tunneling model of the type 
originally proposed by Wheatley, Hsu, and Anderson\cite{jmw}. 
The model consists of
a kinetic energy term $H_0$, which is the same as defined 
in Eq. \ref{h0}, and a local singlet pair tunneling potential 
term $H_I$:
\begin{equation}
H_I = -{\lambda\over 4} \sum_{r,\delta =\hat a ,\hat b} \left(
\hat\Delta_{1, r, \delta}^\dagger \hat\Delta_{2, r+{\hat c\over 2}, \delta} 
+\hat\Delta_{1, r, \delta}^\dagger\hat\Delta_{2, r-{\hat c\over 2}, \delta}
+{\rm h.c.}\right),\label{pairham}
\end{equation}
where $\hat\Delta_{n, r, \delta}\equiv c_{nr\uparrow}c_{nr+\delta\downarrow} 
-c_{nr\downarrow}c_{nr+\delta\uparrow}$ is the singlet pair operator. 
This model is compatible with the two experimental 
requirements specified above.
(1) The pairing functions in the chain and plane layers
must have the same symmetry; i.e. both s or both d. Otherwise, 
the only self-consistent solution 
is the trivial one $\tilde\Delta_n=0$. This model accounts for the
presence of interchain pairing discussed above.
(2) The magnitudes of the gap parameters on chains and planes 
are tied together; plane and chain layers are not independently 
superconducting.

Taking the BCS mean-field approximation, each term 
in (\ref{pairham}) can be decoupled, for example for the
first term in (\ref{pairham}), as $ \hat\Delta_{1, r, \delta}^\dagger 
\hat\Delta_{2, r+{\hat c\over 2}, \delta} 
\simeq  \langle\hat\Delta_{1, r, \delta}^\dagger \rangle
\hat\Delta_{2, r+{\hat c\over 2}, \delta}
+ \langle\hat\Delta_{2, r+{\hat c\over 2}, \delta}\rangle
\hat\Delta_{1, r, \delta}^\dagger 
-\langle\hat\Delta_{1, r, \delta}^\dagger \rangle
\langle\hat\Delta_{2, r+{\hat c\over 2}, \delta}\rangle$, 
where $\langle\hat\Delta\rangle$ is the thermal average of the operator 
$\hat\Delta$. As the system is translationally invariant, 
$\langle\hat\Delta_{n, r, \delta}^\dagger \rangle$ 
should be $r$ independent. 
If we further assume the amplitude of 
$\langle\hat\Delta_{n, r, \delta}^\dagger \rangle$ is $\delta$
independent, then the gap function $\lambda 
\langle\hat\Delta_{n, r, \delta}^\dagger \rangle$ can be written 
as $\lambda \langle\hat\Delta_{n, r, \delta}^\dagger \rangle
=\tilde\Delta_n\alpha_{n\delta}$, here 
$\tilde\Delta_n$ and $\alpha_{n\delta}$ are respectively the 
amplitude and the phase factor of 
$\lambda \langle\hat\Delta_{n, r, \delta}^\dagger \rangle$. 
$\alpha_{n\delta}$ are determined by the pairing symmetries 
of the chain and plane bands. 
With 
$\alpha_{1\delta}=\alpha_{2\delta}=\alpha_\delta$, the mean-field 
decoupled pairing potential $H_I$ can be written as
\begin{eqnarray}
H_I^\prime & = & \sum_k\Bigl\{ (
\tilde\Delta_2\gamma_{k}c^\dagger_{1k\uparrow}
c^\dagger_{1-k\downarrow}
+\tilde\Delta_1\gamma_{k}c^\dagger_{2k\uparrow}
c^\dagger_{2-k\downarrow}+{\rm h.c.})\nonumber\\ 
&&+{2\over\lambda}\tilde\Delta_1\tilde\Delta_2\Bigr\} ,
\end{eqnarray}
where $\gamma_k=\sum_\delta \alpha_{\delta}\cos k_\delta$. 
On a square lattice, the nearest-neighbor singlet paired state 
can have either extended s-wave symmetry, $\gamma_k=\cos k_a +
\cos k_b$, or d-wave symmetry, 
$\gamma_k=\cos k_a -\cos k_b$. 
As only the d-wave pairing state can have gap nodes 
on the planar Fermi surface, we consider only this case. 

At the mean field level the proximity and pair tunneling models 
$H^\prime_I$ and $H_I$ have the same form; the difference 
lies only in how the self-consistent procedure
for the gap function has been applied. Thus the mean-field quasiparticle 
energy spectra are
identical except that $\Delta_1\gamma_{1k}$ and 
$\Delta_2\gamma_{2k}$ in the proximity model spectrum 
should now be replaced by  $\tilde\Delta_2
\gamma_k$ and $\tilde\Delta_1\gamma_k$, respectively.
 
\begin{figure}
\leavevmode
\epsfxsize=8.5cm
\epsfbox{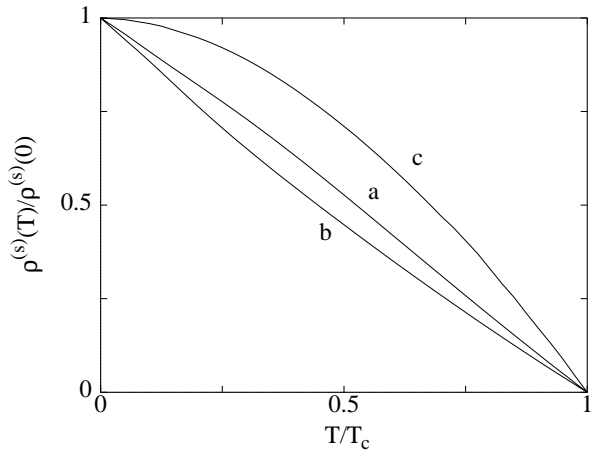}
\caption{Normalized superfluid densities $\rho^{(s)}(T)/ \rho^{(s)}(0)$ 
as functions of $T/T_c$ along $a$, $b$, and $c$ directions in the 
interlayer pair tunneling model with d-wave pairing
symmetry. 
$t=1$, $t_c=0.85$, $t_\perp =0.17$, $\varepsilon_c=0.5$, $\lambda =1.5$ 
($T_c\sim 0.17$), and the electron concentration on the 
CuO planes is about 0.85. 
\label{fig3}}
\end{figure}

Figure \ref{fig3} shows the components of the normalized superfluid 
tensor as functions of $T/T_c$ computed in the pair tunneling 
model using a typical parameter set. 
The ratio $\rho_b^{(s)}/ \rho_a^{(s)}$ at zero temperature 
is mainly determined by $t_c / t$, while 
$\rho_b^{(s)}/\rho_c^{(s)}$ is mainly determined by $t_\perp / t$. 
For the case shown in Figure \ref{fig3}, 
the parameters have been chosen so that 
the ratios $\rho_b^{(s)}/ \rho_a^{(s)}$ 
and $\rho_b^{(s)}/\rho_c^{(s)}$ at zero temperature 
are approximately 2.4 and 100, which gives a qualitative fit to the 
experimental values for $YBa_2Cu_3O_{6.95}$. 
The overall temperature dependences of  
$\rho_b^{(s)}(T)/ \rho_b^{(s)}(0)$ and 
$\rho_a^{(s)}(T)/ \rho_a^{(s)}(0)$ are quite similar. 
At low temperature $\rho_a^{(s)}(T) \sim T$ 
because of the planar d-wave state. 
As both $\tilde\Delta_1$ and $\tilde\Delta_2$ are 
finite in this case, from the previous discussion $\rho 
(\omega) \sim \omega$ at low energy and 
$\rho_b^{(s)}\sim T$ at low temperature. 
The temperature dependence of $\rho_c^{(s)}(T)/ \rho_c^{(s)}(0)$ 
is quite different from $\rho_a^{(s)}(T)/ \rho_a^{(s)} (0)$, 
especially at low temperature where the numerical results 
indicates $\rho_c^{(s)} \sim T^2$.  
All of these results agree very well with the experimental 
measurements\cite{zhang}. 
Increasing $\varepsilon_c$ so that the effective electron 
concentration in the 
chain band decreases, we find that the difference between 
$\rho_a^{(s)}(T)/ \rho_a^{(s)}(0)$ and
$\rho_b^{(s)}(T)/ \rho_b^{(s)}(0)$ becomes smaller, 
but the difference between $\rho_c^{(s)}(T)/ \rho_c^{(s)}(0)$ and 
$\rho_a^{(s)}(T)/ \rho_a^{(s)}(0)$ becomes even larger. 
These changes of $\rho_a^{(s)}(T)/ \rho_a^{(s)}(0)$, 
$\rho_b^{(s)}(T)/ \rho_b^{(s)}(0)$, and $\rho_c^{(s)}(T)/ \rho_c^{(s)}(0)$
with the electron concentration in the chain band agree qualitatively with 
the experimental results for $YBa_2Cu_3O_{6.6}$ and 
$YBa_2Cu_3O_{6.95}$\cite{zhang}. 

In the above discussion we have assumed that the amplitude of the 
order parameter $\langle {\hat \Delta}_{n,\delta}\rangle$ is 
the same in $a$ and $b$ directions. 
A more complete self-consistent 
solution of the gap equations should allow admixture of an 
s-wave component to form a 
d+s wave state\cite{li}.  The gap nodes survive for weak admixture, 
but the positions of the nodes 
shift away from the diagonals of the Brillouin zone. 
Admixture of a small s-component is 
consistent with the c-axis Josephson tunneling experiments on 
$YBa_2Cu_3O_{7-\delta}$\cite{sun}. 
While full self-consistent treatment for the gap 
equations including a s-component is complicated technically 
in our model, preliminary calculations suggest that  
this improves the agreement between the pair tunneling model and 
experiments. Another refinement of the model 
would include a finite plane layer pairing interaction.

In conclusion, we have pointed out that the planar anisotropy of 
electromagnetic response of clean YBCO-type structures
forms a useful probe of the 
microscopic state of Copper-Oxide superconductors. 
We have argued that the data of Hardy and co-workers\cite{zhang} 
imply that the CuO chains layers are intrinsically superconducting 
but that there is a node on the chain Fermi surface sheet. 
The experimentally observed temperature dependences of superfluid 
densities are incompatible with conventional d-wave proximity models 
in the clean limit without an implausible fine tuning of 
parameters. The data are however readily compatible with a model of 
d-wave pair tunneling superconductivity.

\noindent\underline{\it Acknowledgement:}
We thank J. Waldram for bringing our 
attention to the significance of Ref. \cite{zhang}.

\end{document}